\documentclass[aps,prl,superscriptaddress,twocolumn, showpacs,preprintnumbers,amsmath,amssymb]{revtex4-2}

\usepackage{hyperref}
\hypersetup{ colorlinks,
linkcolor=blue,
filecolor=blue,
urlcolor=blue,
citecolor=blue }

\usepackage{amsmath}
\usepackage{mathtools}
\usepackage{xcolor}
\usepackage{ulem}
\usepackage{xr}

\usepackage{subfigure}
\usepackage[pdftex]{graphicx}

\usepackage{geometry}
\usepackage{color} 
\usepackage[font=small,format=plain,labelfont=bf,up,textfont=it,up]{caption}

\begin{document}

\title{Modelo SIS modificado aplicado em um Apocalipse Zumbi com exterminadores}  

\author{G. V. Sousa}
\affiliation{Grupo de Redes Complexas Aplicadas de Jataí, Universidade Federal de Jataí, Jataí/GO, Brasil.}
\author{U. F. Kaneko}
\affiliation{Laboratório Nacional de Luz Síncrotron (LNLS), Centro Brasileiro para Pesquisa em Energia e Materiais (CNPEM), Campinas, São Paulo 13083-970, Brazil.}
\author{P. F. Gomes}
\affiliation{Grupo de Redes Complexas Aplicadas de Jataí, Universidade Federal de Jataí, Jataí/GO, Brasil.}

\begin{abstract}
Neste trabalho estudamos a dinâmica de um apocalipse no qual parte da população se torna zumbis temporariamente, podendo se tornar vivo novamente ou morrer. Descrevemos essa dinâmica utilizando uma versão modificada do modelo epidêmico SIS (do inglês: \textit{susceptible-infected-susceptible}). Os zumbis podem morrer quando em contato com um indivíduo vivo do tipo exterminador. Para definir quem interage com quem utilizamos a rede aleatória Erdös-Rényi e calculamos como as fases absorvente e ativa do modelo SIS original se altera em função do grau médio da rede. Além disso a densidade de mortos atua como uma força dissipativa diminuindo a população efetiva.
\end{abstract}

\maketitle

\section{INTRODUÇÃO}

Os chamados modelos epidêmicos referem-se a um conjunto de dinâmicas onde um indivíduo saudável pode se tornar infectado quando em contato com outros indivíduos infectados. Além desses dois estados, há vários outros que podem ser incluídos no modelo de forma a deixar a dinâmica mais sofisticada como: morto, recuperado, latente, exposto, etc \citep{Grassberger1983,Hethcote2000}. Modelos epidêmicos tem sido amplamente estudados em diversas aplicações nas suas diversas variações \citep{Barthelemy2005,Silva2015,Miranda2020,Reia2022}. Estes modelos apresentam uma fase absorvente a partir da qual o sistema não consegui sair, como também uma fase ativa onde as densidades dos parâmetros de ordme não variam com o tempo \cite{Hinrichsen2000,Vilela2020}. Neste trabalho utilizamos o modelo epidêmico SIS para descrever a dinâmica entre indivíduos vivos e zumbis em um apocalipse zumbi.

\section{FUNDAMENTAÇÃO TEÓRICA}

\subsection{Modelo SIS} \label{sismodel}

O modelo epidêmico SIS considera apenas dois estados para os indivíduos: saudável (susceptible) e infectado (infected). Um indivíduo saudável pode se tornar infectado quando em contato com outros indivíduos infectados. Já um indivíduo infectado pode se tornar saudável com uma chance constante ao longo do tempo, sem a necessidade de ter contato com indivíduos vivos. Esse modelo descreve bem por exemplo uma gripe comum. 

Neste trabalho usaremos os estados vivo e zumbi no lugar de saudável e infectado. Seja $N$ o número total de indivíduos do sistema dos quais $V$ refere-se ao número de indivíduos vivos e $Z$ o número de zumbis. As respectivas densidades serão $v = V/N$ e $z=Z/N$. Neste trabalho iremos analisar o sistema no chamado estado estacionário, no qual as densidades não se alteram mais com o tempo. Este estado é obtido após um certo intervalo de tempo. 

A equação diferencial que descreve o modelo SIS no estado estacionário é \citep{Newman2010}:
\begin{equation}
    \frac{dv}{dt} = c z - bvz, \qquad \qquad \frac{dz}{dt} = bvz - c z, \nonumber
\end{equation}
onde $b$ e $c$ são as chances de conversões de vivo para zumbi e vice-versa. Estas equações referem-se a aproximação de campo médio na qual cada indivíduo interage com todos os outros. Veja que se $z=0$ os vivos continuam vivos, já os zumbis sempre se tornam zumbis após algum tempo, independente da existência de indivíduos vivos. Como temos que $v+z = 1.0$, podemos combinar essas duas equações em uma:
\begin{equation}
    \frac{dz}{dt} = (b - cb - bz)z. \label{slkejriuuuu}
\end{equation}
A solução é:
\begin{equation}
    z(t) = (1- c/b) \frac{De^{(b-c)t}}{1+De^{(b-c)t}}, \label{solcampomedio}
\end{equation}
onde $D = bz_0/(b-c-bz_0)$ e $z_0$ é a concentração inicial de zumbis. Se $b>c$ temos a chamada fase ativa na qual há uma conversão de vivo para zumbi e vice-versa. Além disso a taxa dessas duas conversões é idêntica de forma que as densidades não mais se alteram com o tempo (havendo apenas a flutuação estatística). Apesar de que o conjunto de indivíduos vivos e zumbi estão se alterando continuamente. 

Por outro lado se $b < c$, a taxa de conversão de zumbi para vivo é maior de forma que o número de zumbis cai para zero e todos os indivíduos se mantém vivos. Esta é a chamada fase absorvente pois uma vez havendo apenas vivos não há chance de alguém se tornar zumbi (é necessário contato com algum zumbi para um vivo se tornar zumbi). Uma vez nessa fase, o sistema não mais sai dela, daí o nome de fase absorvente. A condição $b=c$ é exatamente a transição entre essas duas fases.

\subsection{Rede Erdös-Rényi e domínios}

A dinâmica é definida pelo modelo em questão: modelo epidêmico SIS neste trabalho. Ela define o que acontece quando um indivíduo interage com outro. Também é necessário definir o padrão de comunicação entre os indivíduos: quem interage com quem. Todos interagirem com todos é a suposição mais simples possível, chamada de hipótese de mistura homogênea (do inglês \textit{homogeneous mixing hypothesis}) \cite{Barthelemy2005}. Em casos mais realistas essa hipótese não ocorre, sendo necessário a definição de algum tipo de conectividade entre os agentes. Esta conectividade pode ser alterada por diferentes critérios considerando a distância entre os indivíduos em um plano cartesiano em duas dimensões \cite{Juhasz2015}. Ou pode ser definida utilizando uma rede (ou grafo). 

\begin{figure*}[!htbp] 
\vspace{-2pt}
\begin{center}
\includegraphics[scale = 0.5]{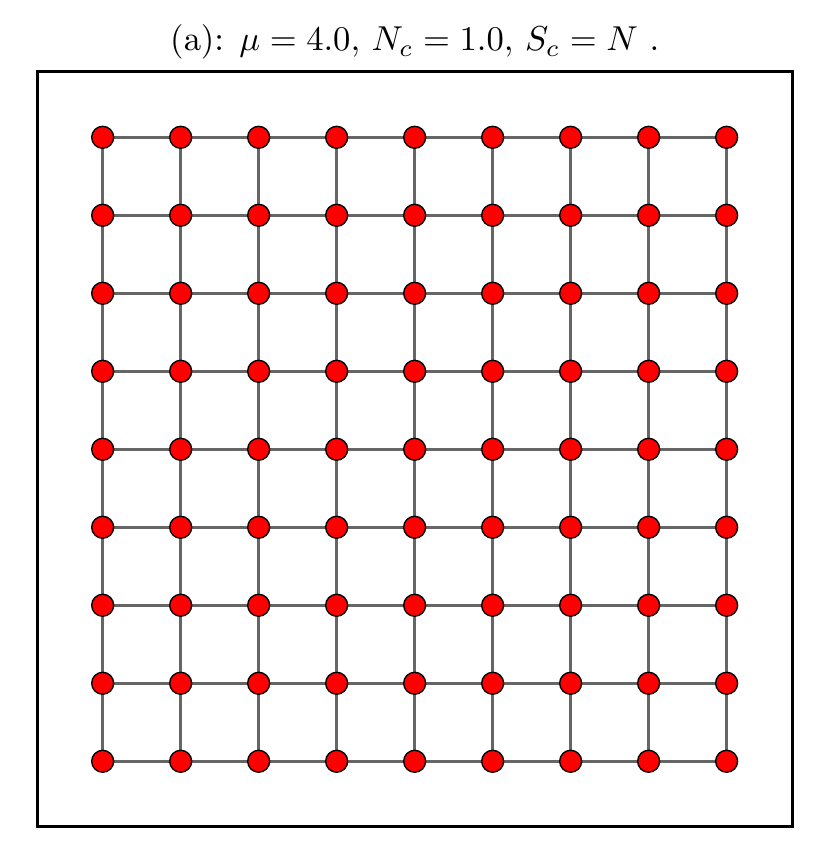}
\includegraphics[scale = 0.5]{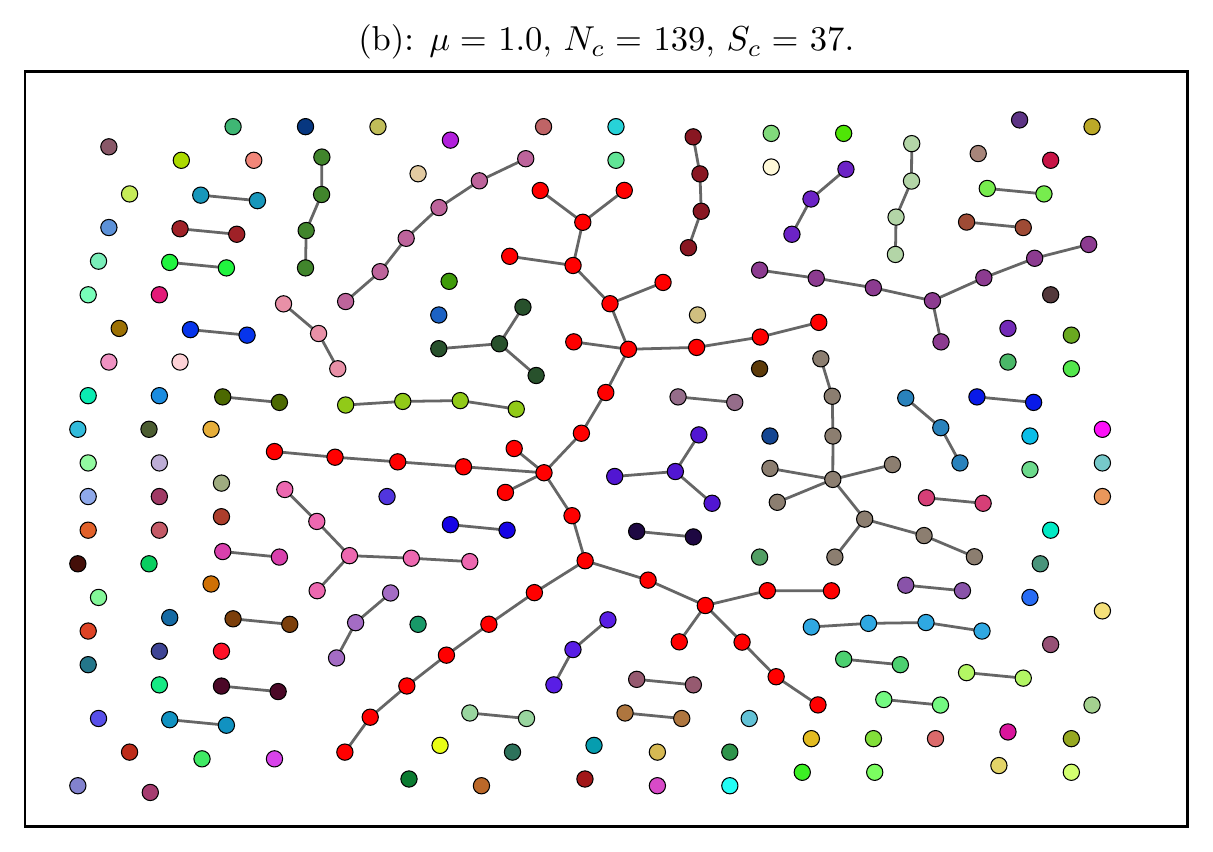}
\includegraphics[scale = 0.5]{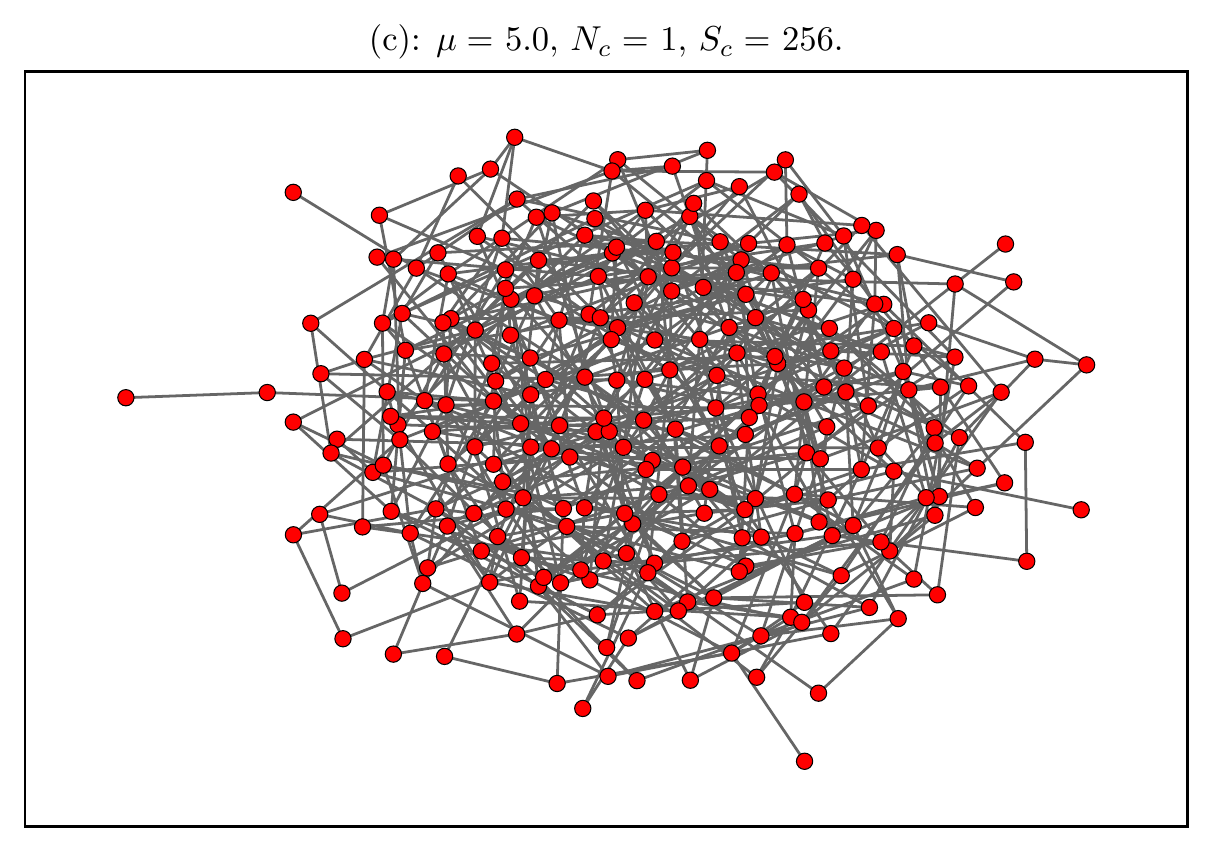}
\caption{(a) Topologia da rede quadrada:  $\mu = 4.0$, $N_c=1$ e $S_c=N$. (b) e (c) Exemplos aletórios da topologia da rede Erdos-Renyi com $N=16^2$. As maiores componentes ficam no centro enquanto os vértices isolados ficam na parte externa. Os vértices da maior componente estão em vermelho para facilitar a visualização. (b) $\mu = 1.0$, $N_c=139$ e $S_c=37$. (c) $\mu = 5.0$, $N_c=1$ e $S_c=N$.}
\label{figura1}
\end{center}
\end{figure*}

Tradicionalmente a rede quadrada (veja figura \ref{figura1}(a)) tem sido utilizada em simulações numéricas de Monte Carlo devido a sua fácil implementação computacional \cite{Silva2015}. Uma rede que utiliza o critério espacial em sua definição e tem sido usada em Dinâmica Social é a chamada rede aleatória geométrica, ou RGG na sigla em inglês \cite{Gomes2019,Vilela2020}. Neste trabalho escolhemos utilizar uma rede aleatória que permite alterar sua topologia: a chamada rede aleatória do tipo Erdös-Rényi (RER) \citep{Solomonoff1561,Erdos1959,Erdos1960}. Sua definição é: cada dupla de vértices tem uma chance $p$ de possuir uma aresta (ou seja, de ser conectada). Os parâmetros $N$ e $p$ são os chamados parâmetros de controle da rede. O grau $k_i$ de um vértice $i$ é o número de conexões (arestas) que ele tem com os outros vértices. Já o grau médio da rede é a média de $k_i$ entre todos os vértices da rede: $\mu = (1/N)\sum_i^N k_i$. No caso da RER temos que $\mu = p(N-1)$ \citep{Barabasi2016}, de forma que podemos utilizar $\mu$ ao invés de $p$ como o parâmetro de controle, por ter um significado mais intuitivo. Uma vez definido $\mu$, basta fazer $p = \mu / (N-1)$ no algoritmo.



\begin{figure}[!htbp] 
\vspace{-2pt}
\begin{center}
\includegraphics[scale = 0.5] {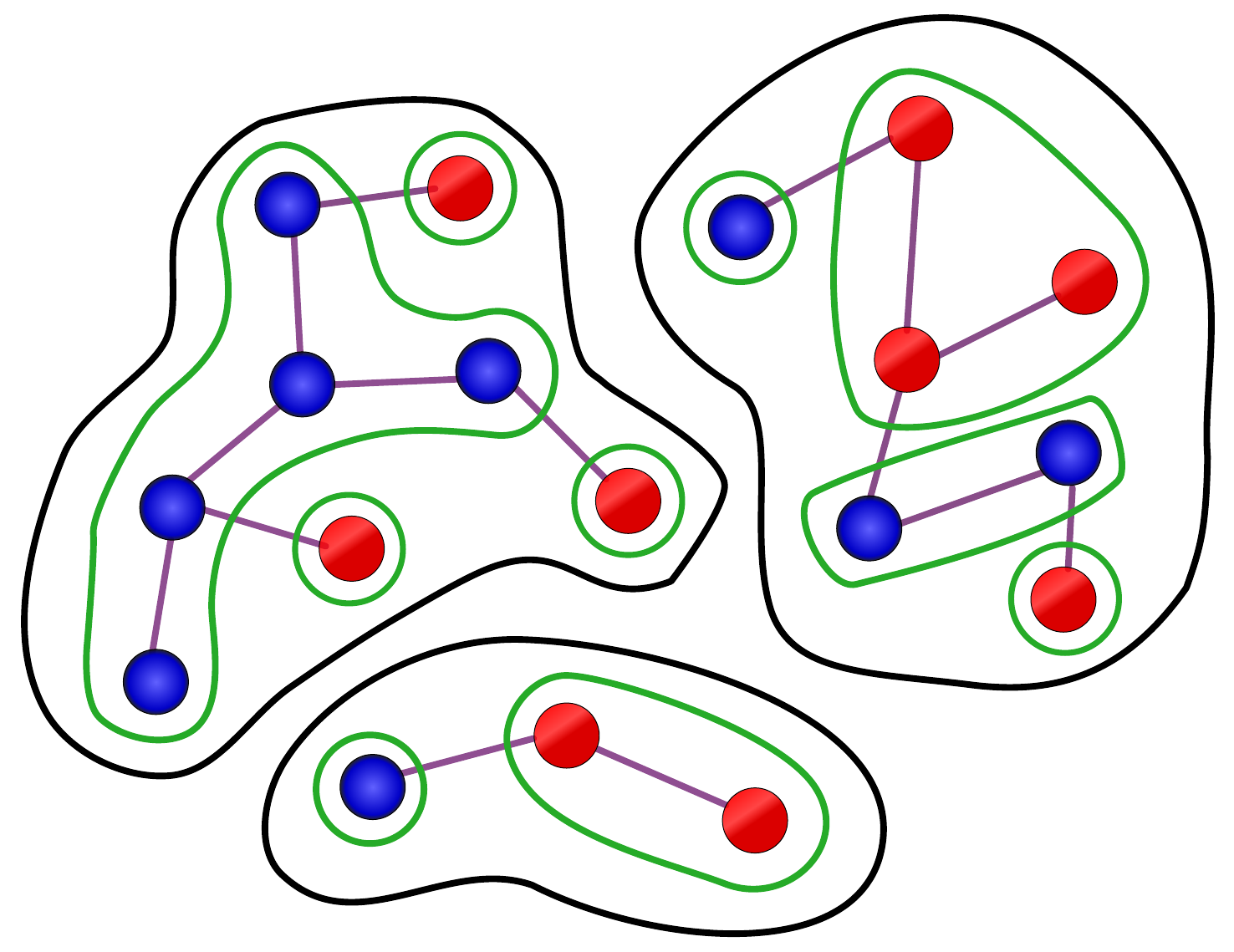}
\caption{Ilustração de componentes e domínio. Os contornos pretos indicam componentes e os contornos verdes indicam domínios. A cor dos vértices (azul ou vermelho) indica o estado deles. Nesta rede temos: $N = 18$, $N_c = 3$, $S_c = 8$, $N_d=10$ e $S_d=5$.}
\label{figura0}
\end{center}
\end{figure}

Uma segunda característica importante de redes é a sua distribuição de componentes. Uma componente é um aglomerado conectado de vértices, e o seu tamanho é o número de vértices contidos \cite{Newman2010}. Um vértice isolado é uma componente de tamanho 1. Seja $N_c$ e $S_c$ o número de componentes e o tamanho (número de vértices) da maior componente de uma rede. Uma rede conectada tem apenas uma componente, logo: $N_c = 1$ e $S_c=N$. Além das componentes, podemos definir também os domínios: grupo de vértices conectados dentro de uma componente tendo o mesmo estado \cite{Reia2019a,Reia2020a,Gomes2022}. Seja $N_d$ e $S_d$ o número de domínios e o tamanho do maior domínio de uma dada rede. Este conceito é apenas aplicado em redes nas quais os vértices tem estados diferentes, como em disseminação cultural \cite{Reia2016,Axelrod1997,Saberi2015,Klemm2005} e modelos epidêmicos. A figura \ref{figura0} mostra uma ilustração desses conceitos em uma rede fictícia. Nas análise utilizamos as densidades desses parâmetros:
\begin{eqnarray}
n_c &=& \frac{N_c}{N}, \qquad \qquad s_c = \frac{S_c}{N}, \nonumber \\
n_d &=& \frac{N_d}{N}, \qquad \qquad s_d = \frac{S_d}{N}. \nonumber
\end{eqnarray}

\begin{figure}[!htbp] 
\vspace{-2pt}
\begin{center}
\includegraphics[scale = 0.1] {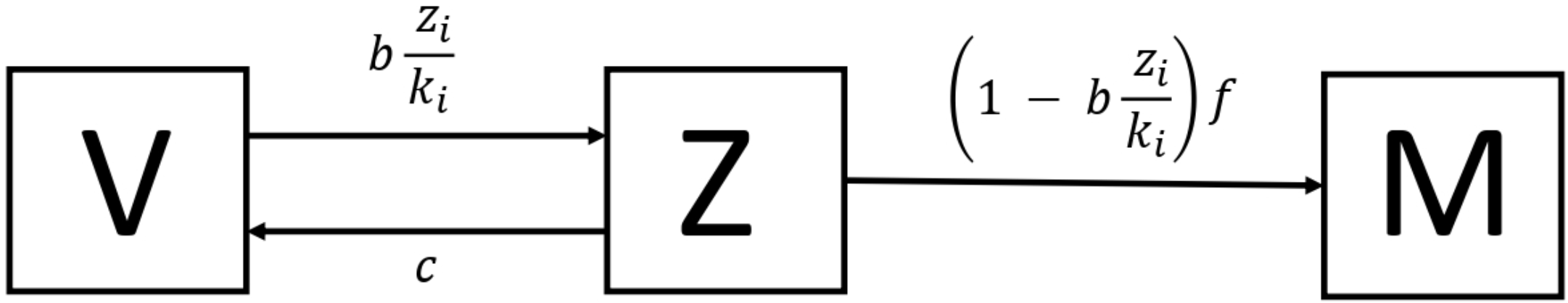}
\caption{Ilustração das mudanças de estados e suas probabilidades.}
\label{figura0a}
\end{center}
\end{figure}

\section{METODOLOGIA}


\begin{table}
\centering
 \begin{tabular}{c|c}    
 \hline
 \hline
 Símbolo & Definição \tabularnewline
 \hline
 \hline $b$  & taxa de infecção \tabularnewline
 \hline  $c$  & taxa de recuperação dos zumbis \tabularnewline
 \hline $f$  & fração de exterminadores \tabularnewline
 \hline  $k_i$ & grau do indivíduo $i$ \tabularnewline
 \hline  $m$ & densidade de mortos \tabularnewline
 \hline  $\mu$  & grau médio da RER \tabularnewline
 \hline  $N$  & número total de indivíduos  \tabularnewline
  \hline $N_c$ & número de componentes \tabularnewline
 \hline $n_d$ & densidade do número de componentes \tabularnewline
  \hline $N_d$ & número de domínios \tabularnewline
 \hline $n_d$ & densidade do número de domínios \tabularnewline
 \hline $p$  & chance de conexão da RER \tabularnewline
 \hline  $Q$ & tempo para atingir o estado estacionário \tabularnewline
 \hline $S_1$ e $S_2$ & número de amostras \tabularnewline
  \hline $S_c$ & tamanho da maior componente \tabularnewline
 \hline $s_c$ & densidade do tamanho da maior componente \tabularnewline
 \hline $S_d$ & tamanho do maior domínio \tabularnewline
 \hline $s_d$ & densidade do tamanho do maior domínio \tabularnewline
 \hline $v$ & densidade de vivos \tabularnewline
 \hline $z$ & densidade de zumbis \tabularnewline
 \hline $z_0$ & $z(t)$ no instante inicial \tabularnewline
 \hline 
 \hline 
 \end{tabular}
\caption{Definição de todos os parâmetros utilizados no modelo.}
 \label{tabelaparametros}
 \end{table}

Nosso modelo é uma adaptação do modelo SIS e utilizamos os estados vivo e zumbi no lugar dos estados saudável e infectado. A dinâmica entre esses estados é exatamente a do modelo SIS descrita acima. Todos os parâmetros estão definidos na Tabela \ref{tabelaparametros} para facilitar o entendimento. A seguir descrevemos os passos de uma análise, na qual $\Gamma$ é um número aleatório entre 0.0 e 1.0.
\begin{itemize}
\item Um indivíduo $i$ aleatório é selecionado: $i=1+I(\Gamma N)$, inde $I(a)$ é a função inteiro, que retorna a parte inteira de $a$.
    \item Se o indivíduo selecionado é zumbi, ele pode se tornar vivo com uma chance $c$. Se $\Gamma < c$, esse indivíduo passa de zumbi para vivo.
    \item Se o indivíduo $i$ selecionado é vivo, ele pode se tornar um zumbi com chance $bz_i/k_i$, onde $b$ é a taxa de infecção, $z_i$ é o número de vizinhos zumbi e $k_i$ é o número total de vizinhos do indivíduo $i$. Se $\Gamma < bz_i/k_i$, o indivíduo passa de vivo para zumbi.
    \item Se $\Gamma > bz_i/k_i$ e esse indivíduo vivo também for exterminador, um vizinho zumbi aleatório passa para o estado morto. 
\end{itemize}
Os 3 passos anteriores constituem uma análise.  Uma ilustração dessas conversões está na figura \ref{figura0a}. Se não houver zumbis no sistema, ou se o indivíduo vivo selecionado para a análise não tiver vizinhos zumbis, nada acontece na análise (mas ela é contabilizada). Um passo de Monte Carlo é definido como $N$ análises. Na média todos os indivíduos são analisados, mas em um dado passo pode haver indivíduos não analisados e outros analisados mais de uma vez. A termalização constitui de $Q$ passos de Monte Carlo (uma calibração é feita para se obter o valor adequado desse parâmetro).

Os exterminadores perfazem uma fração aleatória $f$ da população de vivos definidos no instante inicial. Quando um vivo exterminador se torna zumbi e eventualmente se torna vivo novamente, ele não volta como exterminador. Assim a população de exterminadores diminui com o tempo até desaparecer. Após esse instante a densidade de mortos fica constante no tempo. 

\begin{figure}[!htbp] 
\vspace{-2pt}
\begin{center}
\includegraphics[scale = 0.55]{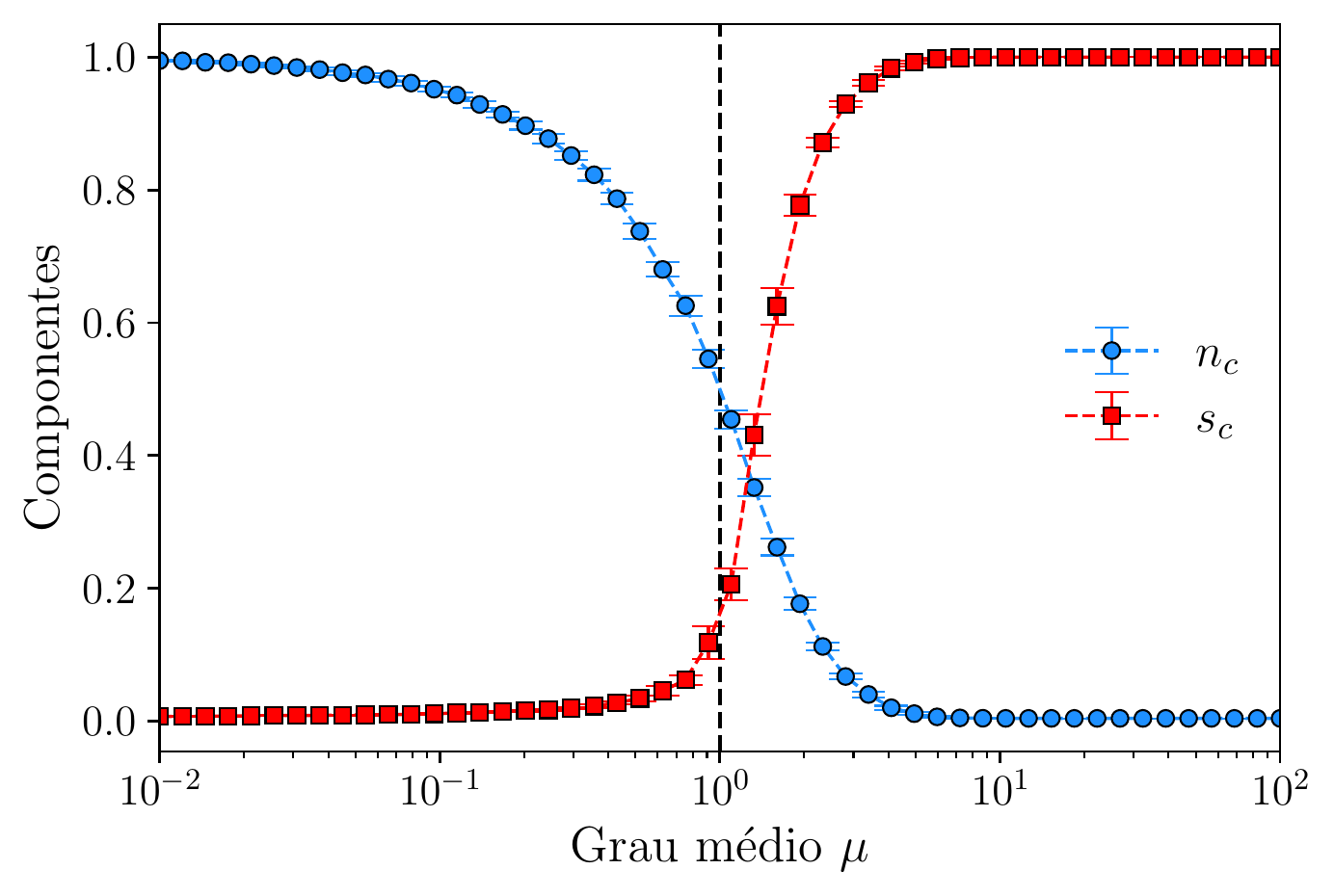}
\caption{Número de componentes $n_c$ e tamanho da maior componente $s_c$ (normalizados) em função do grau médio $\mu$ para a RER. O eixo horizontal está em escala log. A linha vertical tracejada indica a transição de fase em $\mu = \mu_c$. Fase não percolada: $\mu < \mu_c$. Fase percolada: $\mu > \mu_c$.}
\label{figura1d}
\end{center}
\end{figure}


O modelo foi implementado em linguagem Fortran para a geração dos dados enquanto sua análise e todos os gráficos foram feitos em Python. Os seguintes pacotes em Python foram utilizados: Numpy \citep{Harris2020}, Matplotlib \citep{Hunter2007}, NetworkX \citep{Hagberg2008} e Pandas \citep{McKinney2010}. Todos os dados e códigos estão disponíveis mediante solicitação.
  
\section{RESULTADOS}

Na figura \ref{figura1}(a) está uma ilustração de uma rede quadrada (RQ) na qual pode-se ver que cada vértice está ligado com os seus 4 primeiros vizinhos. A RQ é uma rede dita conectada pois há somente uma componente: $N_c=1$. Assim o tamanho da maior (única) componente é exatamente $S_c=N$. Já na figura \ref{figura1}(b) está uma instância aleatória da RER com $\mu=1.0$. Repare que a maior componente tem $S_c=37$ vértices (em vermelho) enquanto que há $N_c=139$ componentes no total (cada vértice isolado é uma componente de tamanho 1.0). A medida que $\mu$ aumenta, a rede se torna mais conectada: $N_c$ diminui e $S_c$ aumenta. Na figura \ref{figura1}(c) está uma instância da RER gerada com $\mu = 5.0$, na qual a rede já está conectada: $N_c=1$ e $S_c=N$. A figura \ref{figura1d} mostra a variação de $n_c = N_c/N$ e $s_c=S_c/N$ em função de $\mu$ ao longo de 5 ordens de grandeza. No limite $\mu$ muito pequeno a rede está fragmentada contendo apenas vértices isolados: $n_c=1.0$ e $s_c = 1/N$. Já no limite de $\mu$ muito grande a rede se torna conectada: $n_c=1/N$ e $s_c = 1$. Assim a rede apresenta as fases não percolada e percolada sendo que a transição é em torno de $\mu = 1.0$ \cite{Newman2010,Vilela2020}. Todos os resultados mostrados a seguir são na rede percolada: $\mu \geq 1.0$

\begin{figure}[!htbp] 
\vspace{-2pt}
\begin{center}
\includegraphics[scale = 0.5]{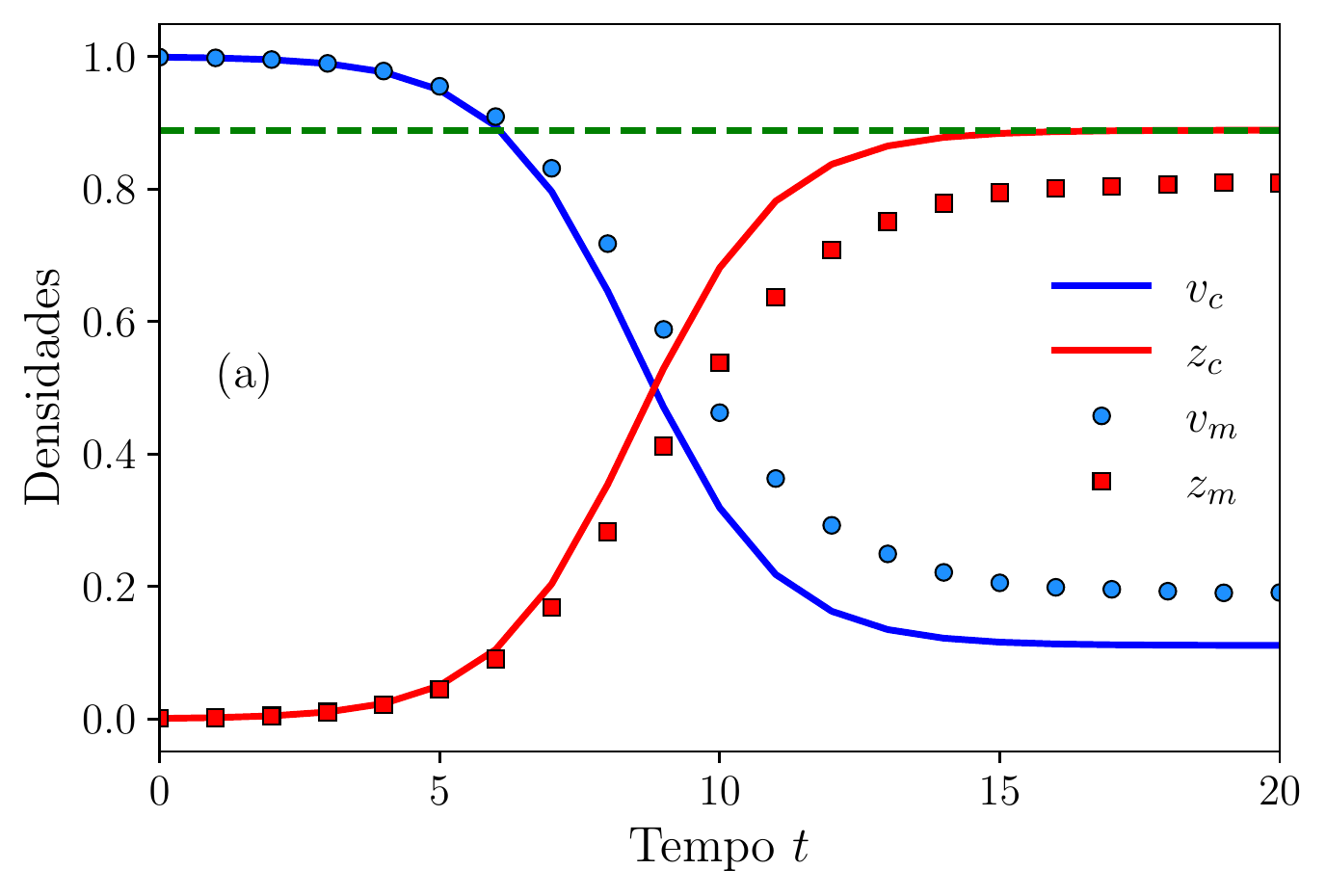}
\includegraphics[scale = 0.5]{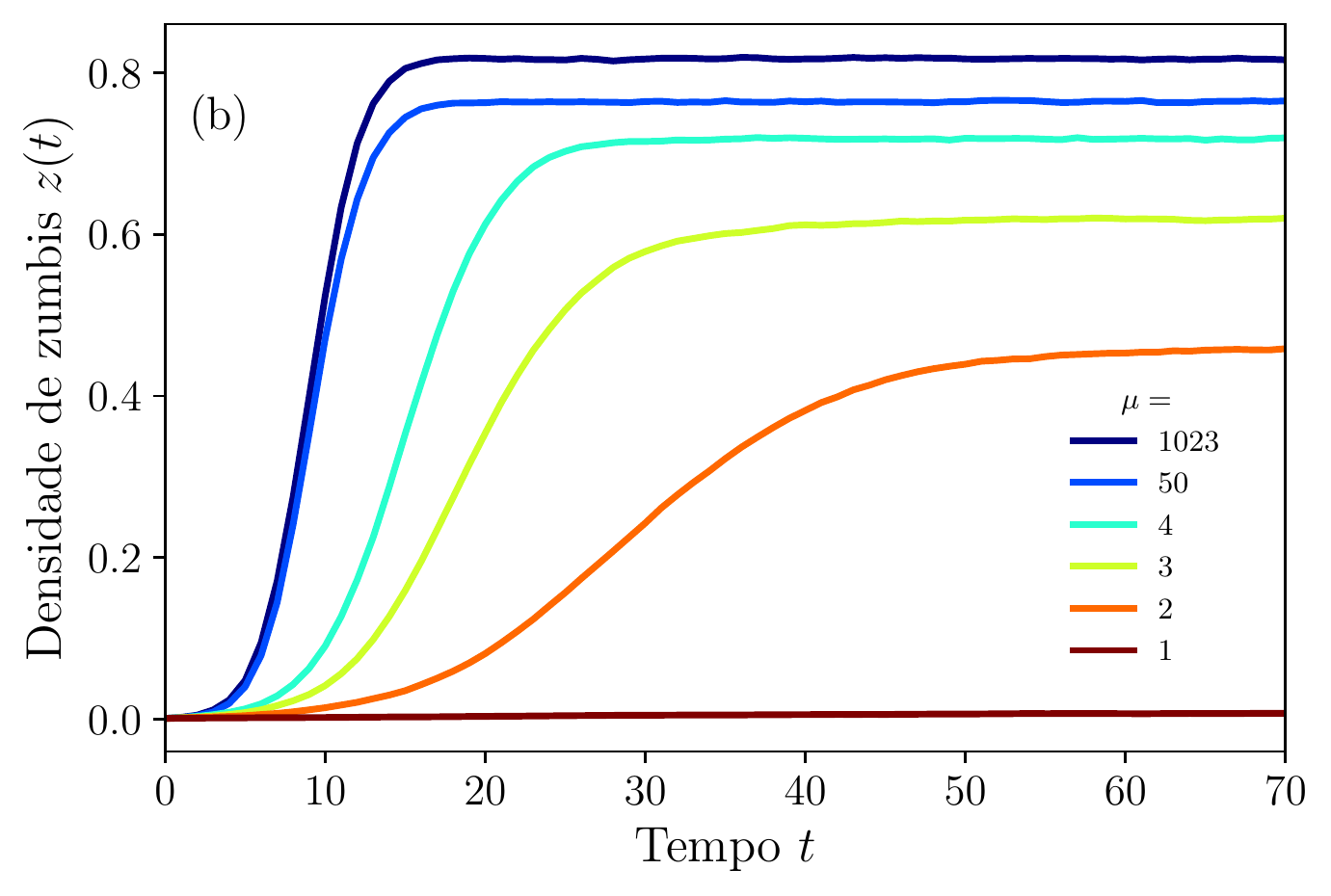}
\caption{Evolução temporal das densidades. Parâmetros: $N=32^2 = 1024$, $b=0.9$, $c=0.1$, $S=100$. A condição inicial para esses resultados é $z_0=z(0)=1/N$, ou seja, no instante inicial havia apenas um zumbi no sistema. (a) Densidades de vivos $v$ e zumbis $z$ vs. tempo de evolução. O sub-índice $c$ indica aproximação de campo médio (Eq. \ref{solcampomedio}) e o sub-índice $m$ Método de Monte Carlo com $\mu = N-1$, o que configura uma rede totalmente conectada. O estado estacionário começa quando as densidades não variam mais com o tempo, neste gráfico, após $t=15$ aproximadamente. As densidades no estado estacionário dependem das taxas $b$ e $c$, sendo a linha tracejada o valor $z_c=(b-c)/b$ na aproximação de campo médio. (b) Densidade de zumbis vs. tempo para diferentes graus médio $\mu$ na RER obtido com Método de Monte Carlo.}
\label{figura2}
\end{center}
\end{figure}

\subsection{Termalização}

Para evidenciar o efeito do uso de uma rede arbitrária, apresentamos uma comparação dos resultados utilizando aproximação de campo médio e Método de Monte Carlo na figura \ref{figura2}(a). Mesmo utilizando $\mu = N-1$ (rede totalmente conectada), o resultado da evolução temporal não é idêntico ao do campo médio. Esta variação temporal também serve para calibrar o tempo mínimo $Q$ para se obter o estado estacionário. Já na figura \ref{figura2}(b) podemos ver como a topologia da rede influencia na evolução temporal da densidade de zumbis. O caso totalmente conectado refere-se a curva $\mu = 1023$, que está bem próxima da curva com $\mu = 50$. Ou seja, para redes densamente conectadas há pouca variação na densidade. Já para $\mu = 4$ e valores menores há uma variação drástica até que em $\mu = 1$ a densidade não apresenta uma evolução significativa. Ou seja, neste caso, como cada indivíduo só tem um vizinho, o zumbi inicial único não foi suficiente para infectar outros zumbis, não houve a formação de um apocalipse.

\begin{figure}[!htbp] 
\vspace{-2pt}
\begin{center}
\includegraphics[scale = 0.5]{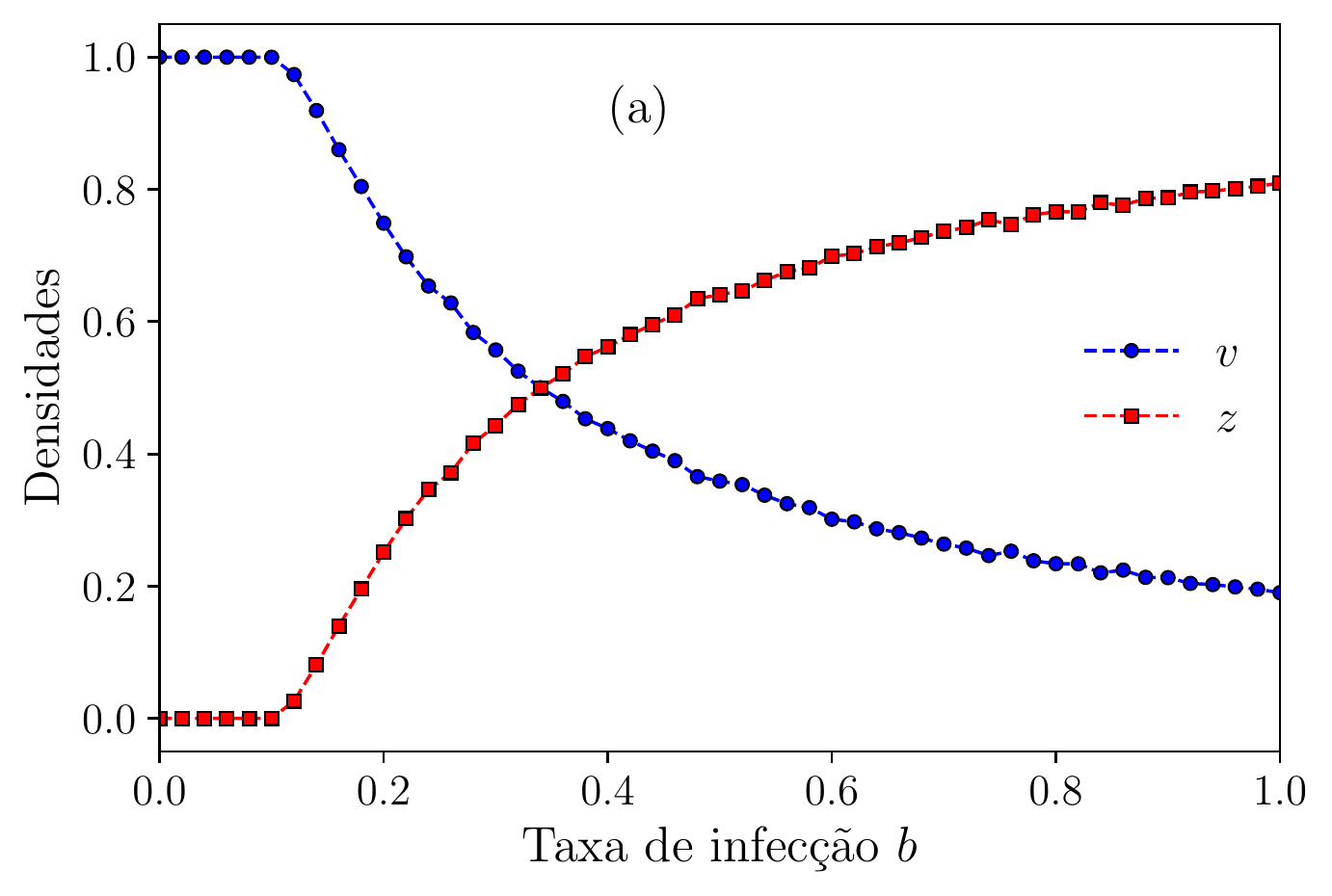}
\includegraphics[scale = 0.5]{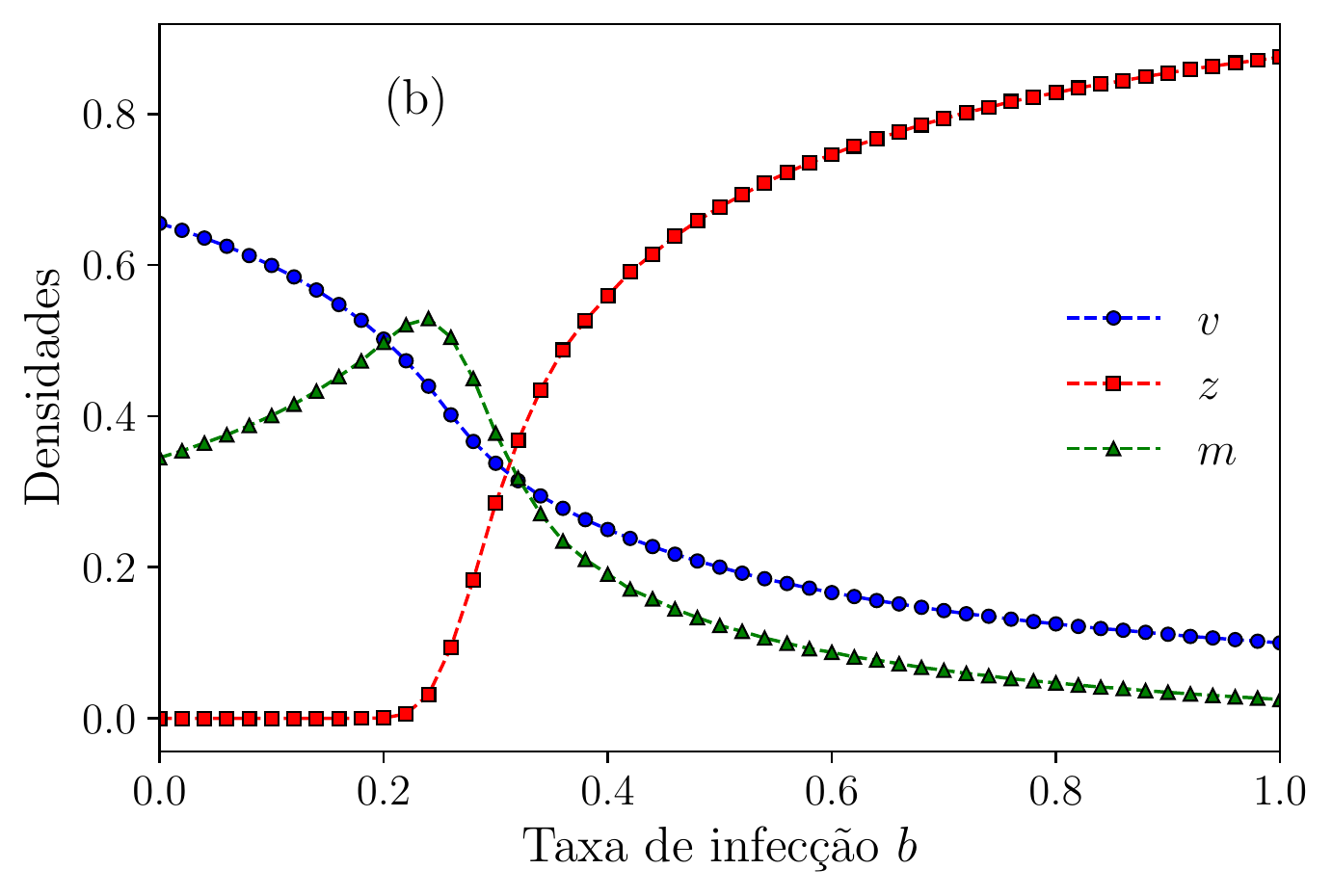} \caption{Densidades em função da taxa de infecção $b$ no estado estacionário considerando uma rede totalmente conectada: $\mu = N-1$. Parâmetros: $N=32^2 = 1024$, $c=0.1$, $S_1=100$ e $Q=500$. Condição inicial: $z_0 = 1/2$, ou seja, metade da população é zumbi no instante inicial $t=0$. (a) Modelo SIS original contendo apenas vivos e zumbis como estados possíveis. $S_2=100$. (b) Modelo SIS modificado contendo os exterminadores e mortos como estados adicionais. $S_2=20$ e $f=0.1$.}
\label{figura3}
\end{center}
\end{figure}

\subsection{Diagrama de fases}

Vamos agora analisar as fases em função do parâmetro de controle $b$. A figura \ref{figura3}(a) apresenta o diagrama de fases do modelo SIS original. A fase absorvente é quando não há zumbis ($z=0$), pois nesse casos os vivos continuam vivos e nada acontece. Essa fase ocorre para valores baixos de $b$. Já para valores maiores a densidade de zumbis é maior que zero e a taxa de conversão de zumbis para vivos é igual a taxa de vivos para zumbi, de forma que ambas as densidades permanecem constantes (a menos da flutuação estatística) em função do tempo. Esta é a fase ativa. De modo geral a transição será em torno de $b=c$. 

A figura \ref{figura3}(b) apresenta o diagrama de fases modificado com a inclusão dos exterminadores (um grupo dos indivíduos vivos). No instante inicial uma fração $f$ aleatória dos $v(0)$ indivíduos vivos é escolhido como exterminadores também. Agora há um terceiro estado possível: mortos (triângulos em verde). A fase absorvente agora é maior sendo que a transição de fase ocorre para $b$ um pouco maior que 0.2. Neste mesmo valor há um máximo da densidade de mortos. A densidade de zumbis tem aproximadamente a mesma forma que no caso SIS original, enquanto a densidade de vivos é bastante alterada

\begin{figure}[!htbp] 
\vspace{-2pt}
\begin{center}
\includegraphics[scale = 0.5]{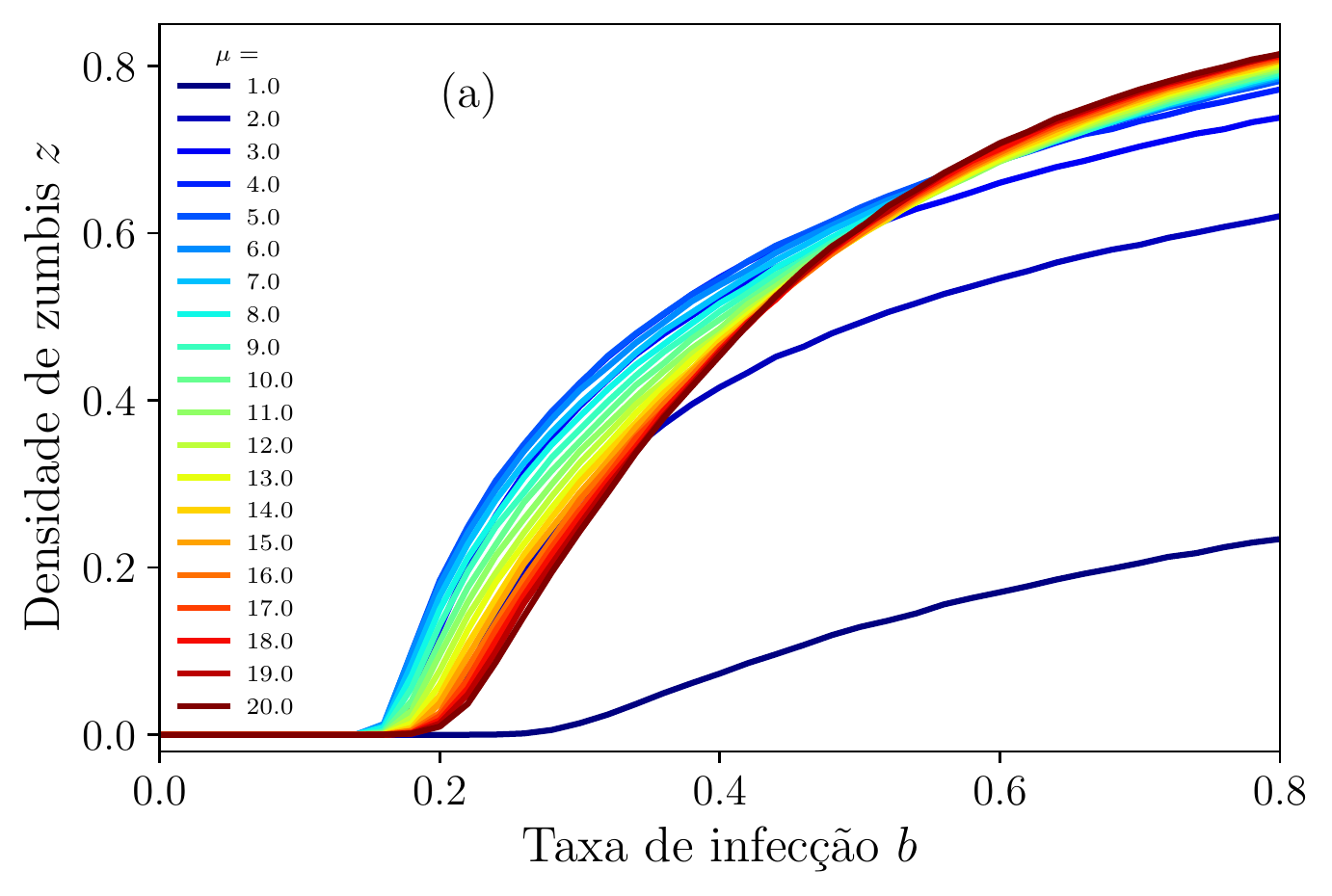}
\includegraphics[scale = 0.5]{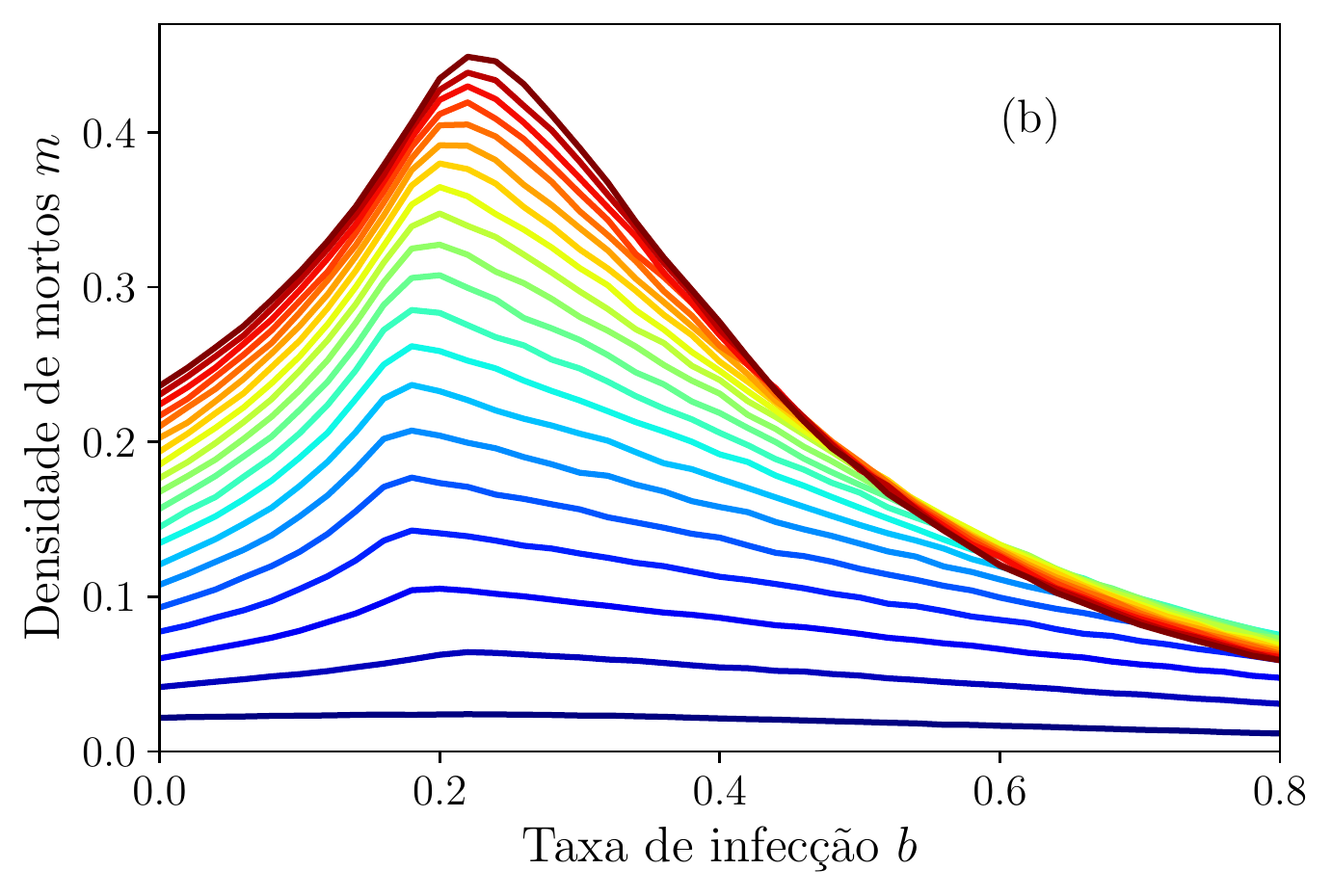}
\includegraphics[scale = 0.5]{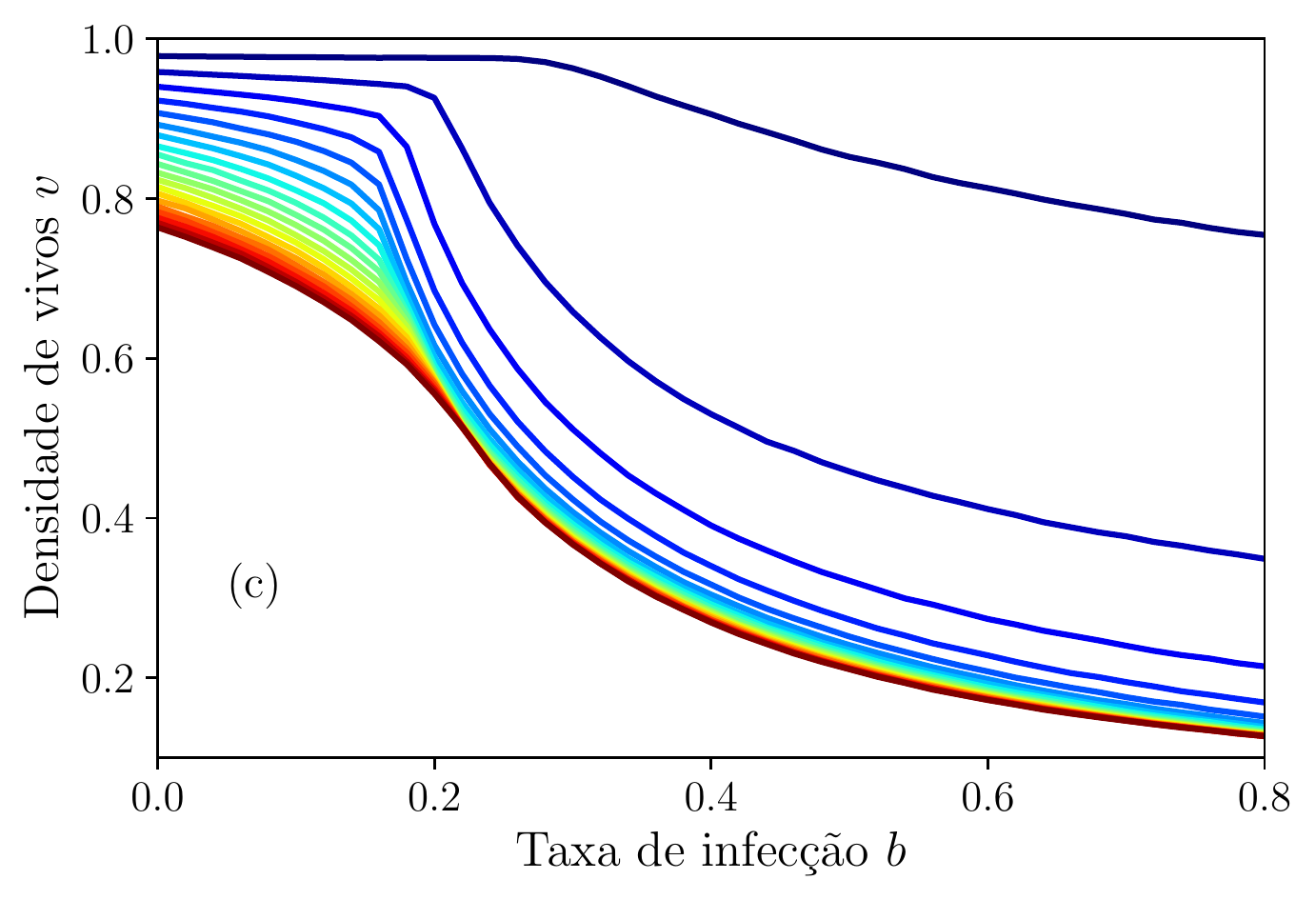}
\caption{Densidades em função da taxa de infecção $b$ para diferentes valores do grau médio $\mu$. Parâmetros: $N=32^2 = 1024$, $c=0.1$, $f=0.1$, $S_1=100$ e $S_2=20$. (a) Densidade de zumbis $z$. (b) Densidade de mortos $m$. (c) Densidade de vivos $v$. A legenda de cores da parte (a) vale para as partes (b) e (c). }
\label{figura4}
\end{center}
\end{figure}

\begin{figure}[!htbp] 
\vspace{-2pt}
\begin{center}
\includegraphics[scale = 0.6]{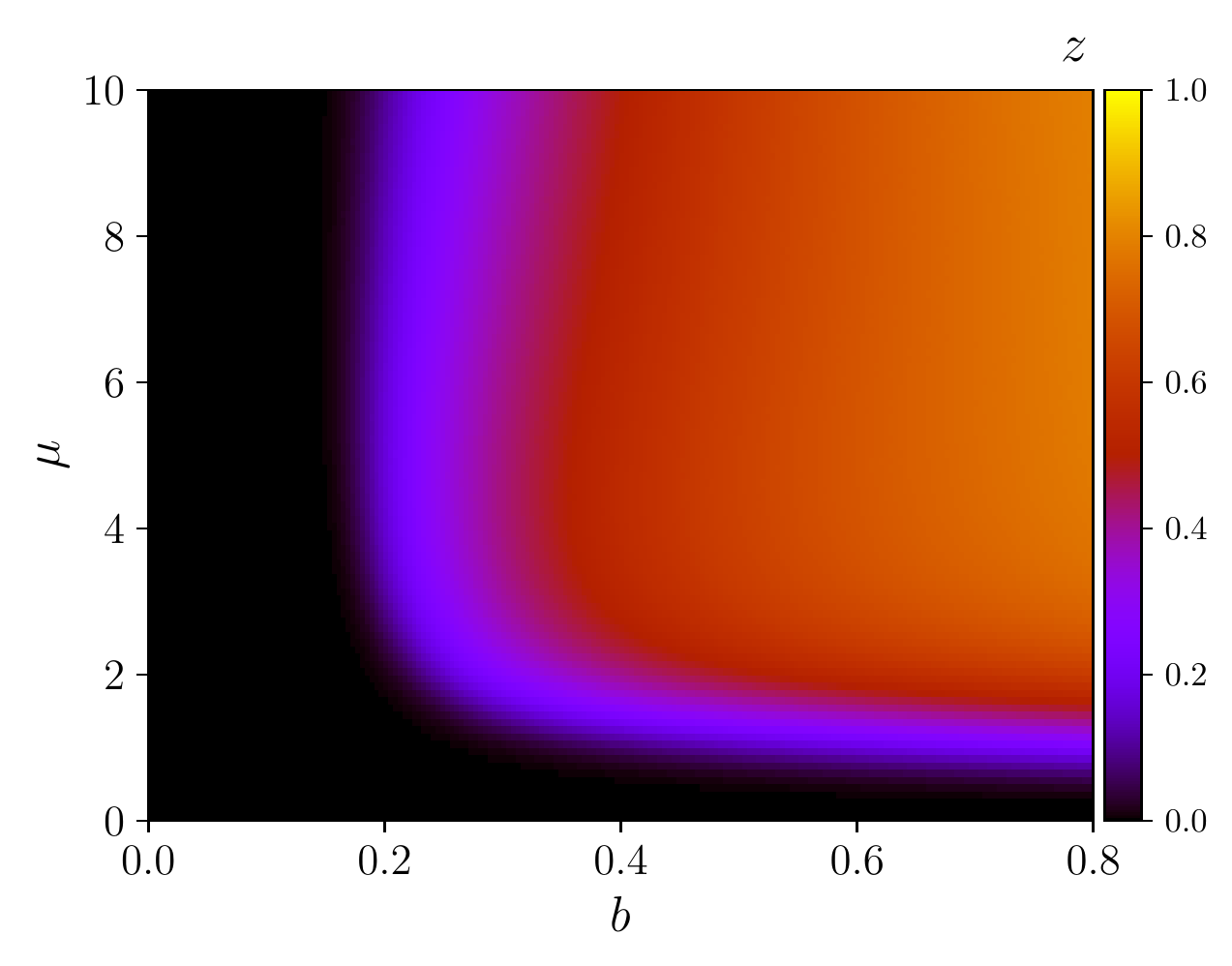}
\caption{Gráfico 2D da densidade de zumbi $z$ no plano $b,\mu$. A fase absorvente ($z=0.0$) é a região preta. Parâmetros: $N=16^2 = 1024$, $c=0.1$, $f=0.1$, $S_1=100$ e $S_2=280$. Os espaçamentos são $\Delta b = 0.004$ no eixo horizontal e $\Delta \mu = 0.1$ no eixo vertical. Q quantidade total de pontos é $201 \times 101 = 20301$.}
\label{figura4d}
\end{center}
\end{figure}

\subsection{Influência da topologia da rede}

Vamos analisar em mais detalhes a influência da topologia da rede, na figura do seu grau médio $\mu$. Como foi observado na figura \ref{figura2}(b) que valores altos de $\mu$ não criam grandes alterações, vamos focar agora em valores mais baixos. Da figura \ref{figura4}(a) podemos observar que o grau médio $\mu$ não altera significativamente a transição de fase. Com exceção de $\mu=1.0$, a transição ocorre para valores de $b$ um pouco antes de 0.2. Lembrando que a fase absorvente ocorre para $z=0$. Já a densidade de mortos na figura \ref{figura4}(b) apresenta um máximo para um $\mu$ maior e um pouco acima de $b=0.2$. Na figura \ref{figura4}(c) podemos observar que a densidade de vivos cai rapidamente na fase ativa a medida que aumenta a taxa $b$. Na figura \ref{figura4d} está o diagrama de fases 2D no plano $(b,\mu)$. Observa-se que a fase absorvente ocorre para $b$ ou $\mu$ pequenos.

\begin{figure}[!htbp] 
\vspace{-2pt}
\begin{center}
\includegraphics[scale = 0.5]{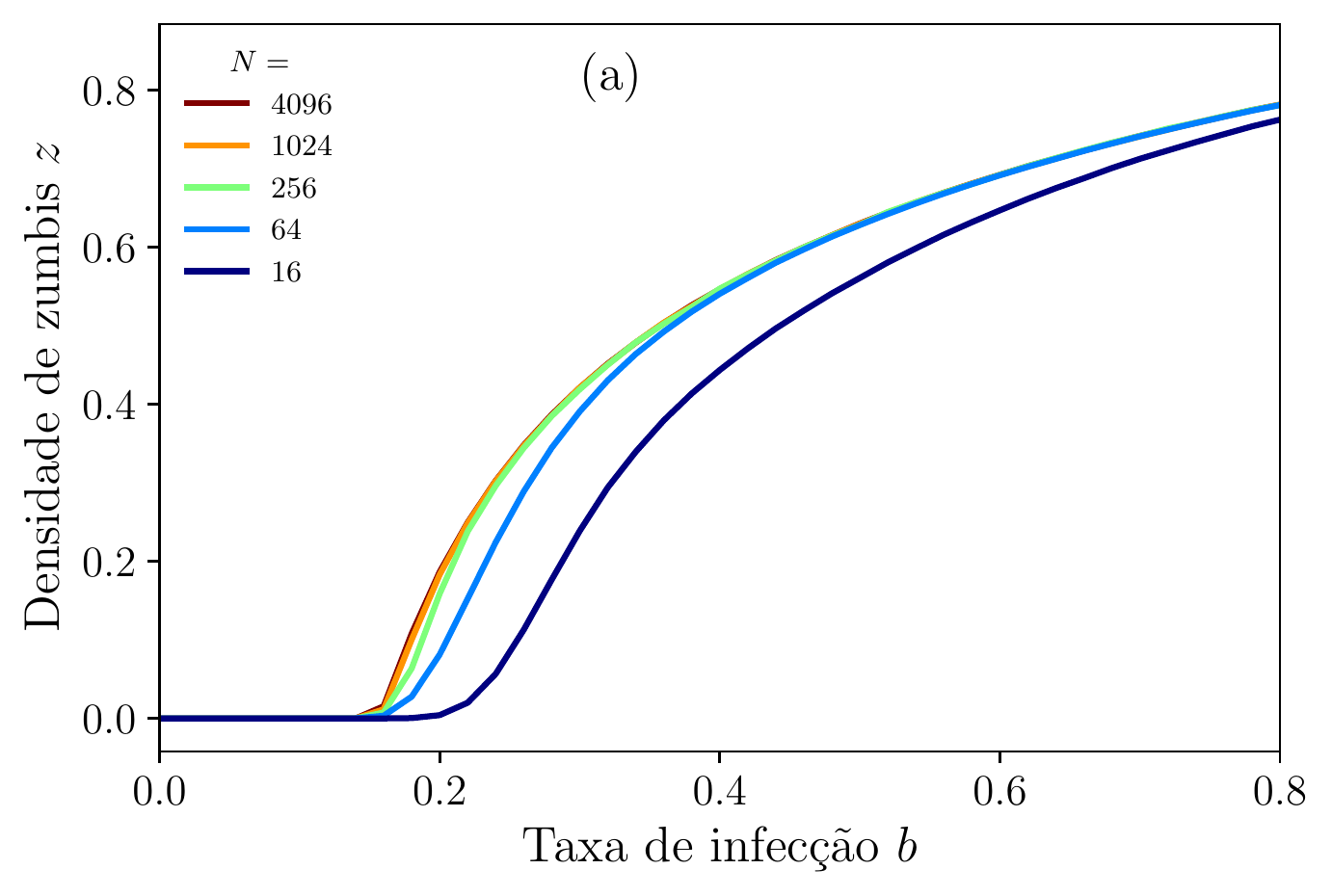}
\includegraphics[scale = 0.5]{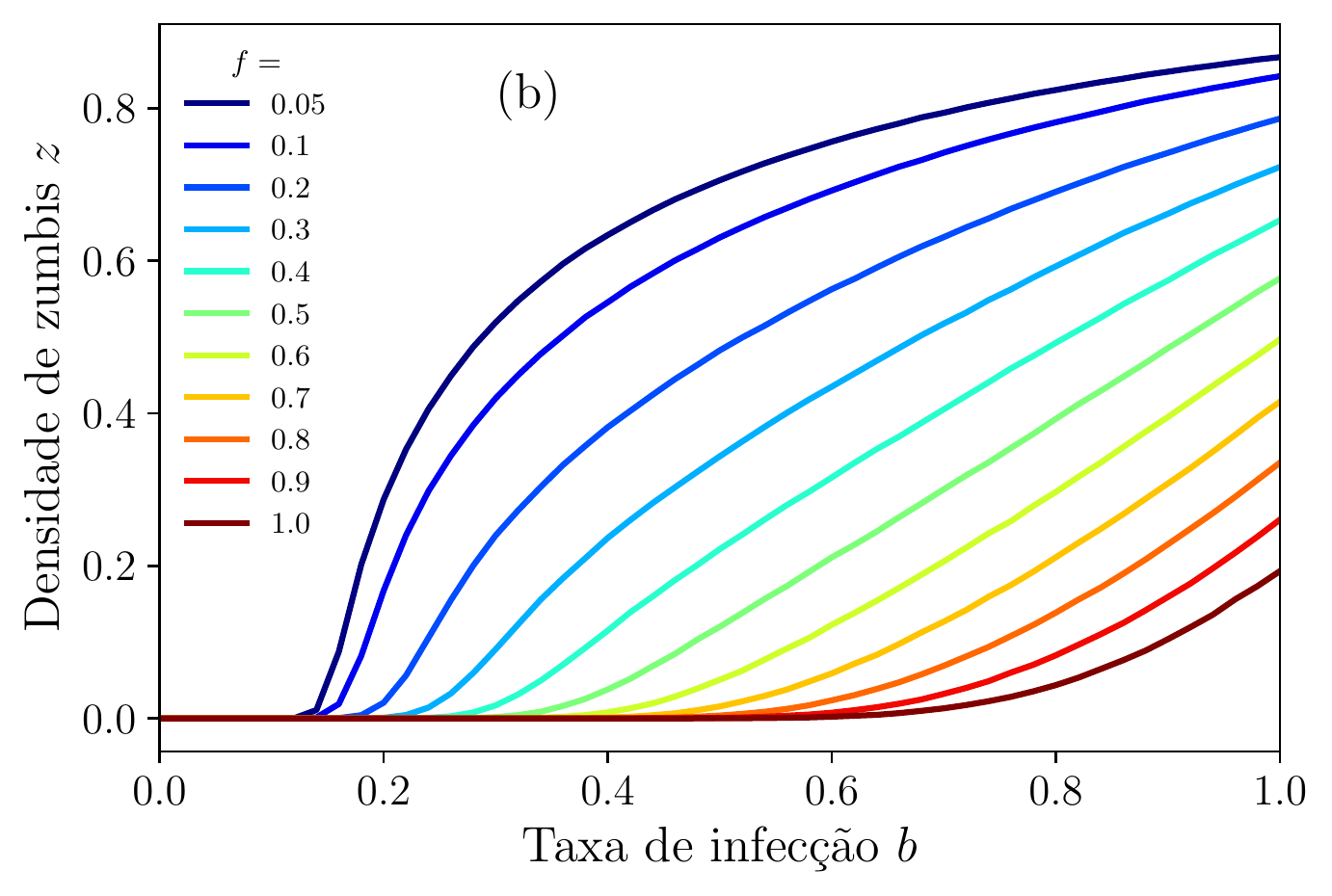}
\includegraphics[scale = 0.5]{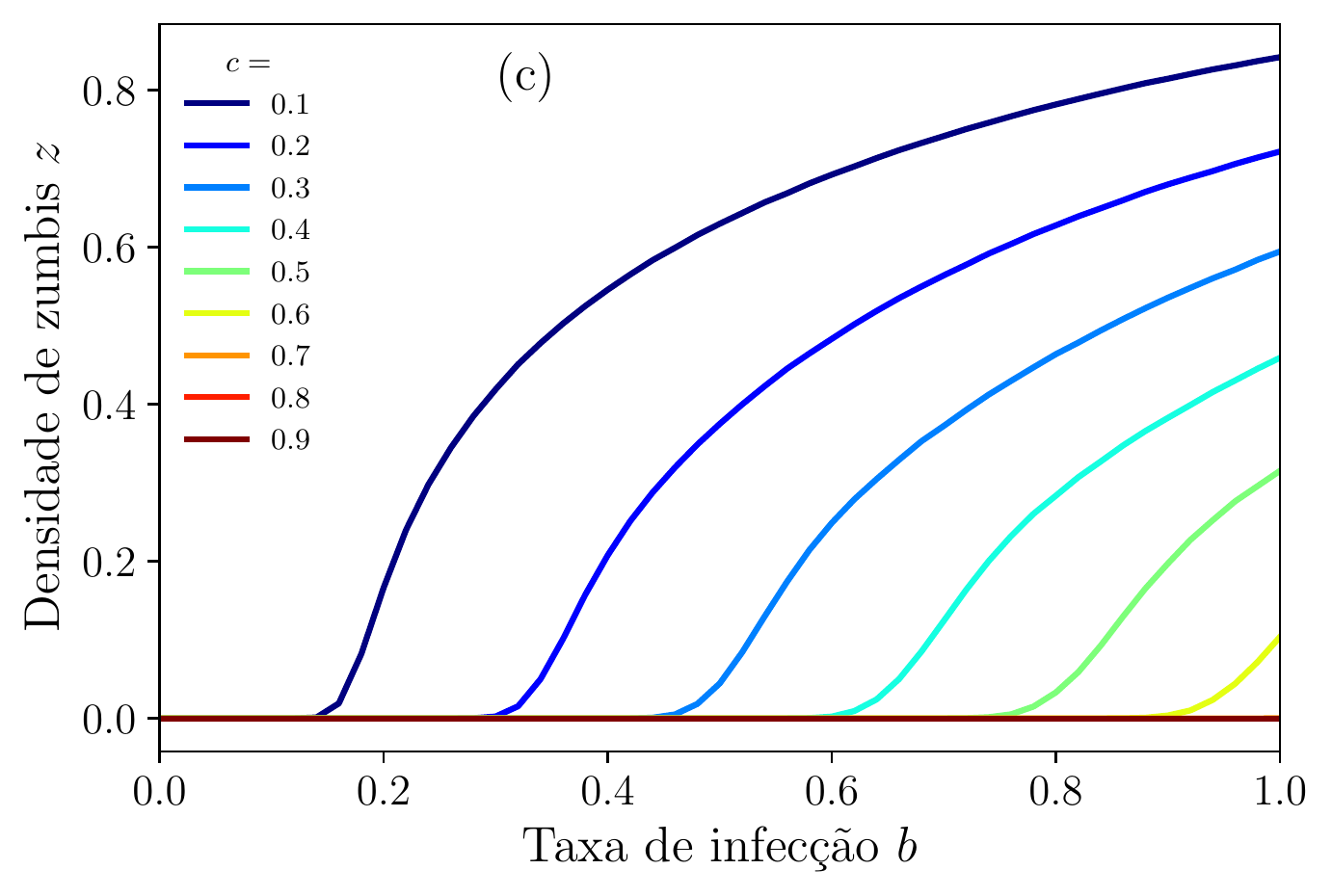}
\caption{Densidades de zumbi em função da taxa de infecção $b$. Parâmetros: $\mu = 5.0$, $Q=500$, $S_1=100$ e $S_2=300$. (a) Variando o tamanho $N$ do sistema. (c) Variando a fração $f$ de exterminadores. (c) Variando a chance de recuperação $c$.}
\label{figura5}
\end{center}
\end{figure}

\subsection{Outros parâmetros}

O diagrama de fase é pensado nos parâmetros $b$ e $\mu$, porém os outros parâmetros do modelo também tem um efeito. Na figura \ref{figura5}(a) está mostrado a influência do tamanho $N$ do sistema. O objetivo é identificar a partir de qual valor a densidade de zumbis não varia mais com $N$. A partir deste resultado observa-se que para $N \geq 16^2$ é adequado. Já a fração $f$ de exterminadores pode inibir a fase ativa, já que um alto valor aumenta a densidade de mortos e diminui a de zumbis (veja figura \ref{figura5}(b)). De fato para $f=1.0$ todos os vivos são exterminadores também e quando analisados eles ou se tornam um zumbi ou matam um deles. Já na figura \ref{figura5}(c) apresentamos a influência da taxa $c$, a chance de um zumbi voltar a ser vivo. O efeito desse parâmetro é semelhante ao de $f$: ambos inibem a fase ativa.

\begin{figure}[!htbp] 
\vspace{-2pt}
\begin{center}
\includegraphics[scale = 0.5]{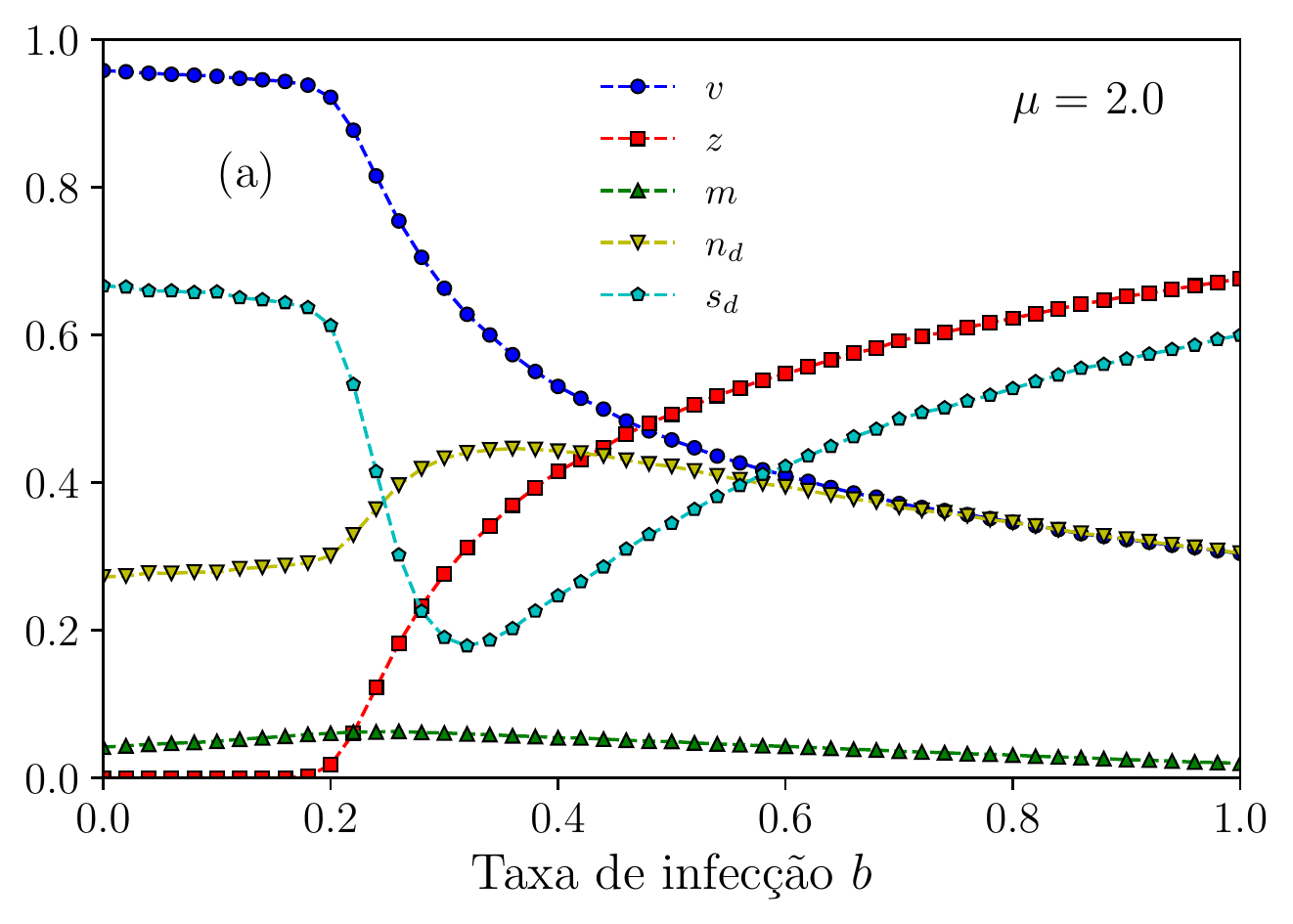}
\includegraphics[scale = 0.5]{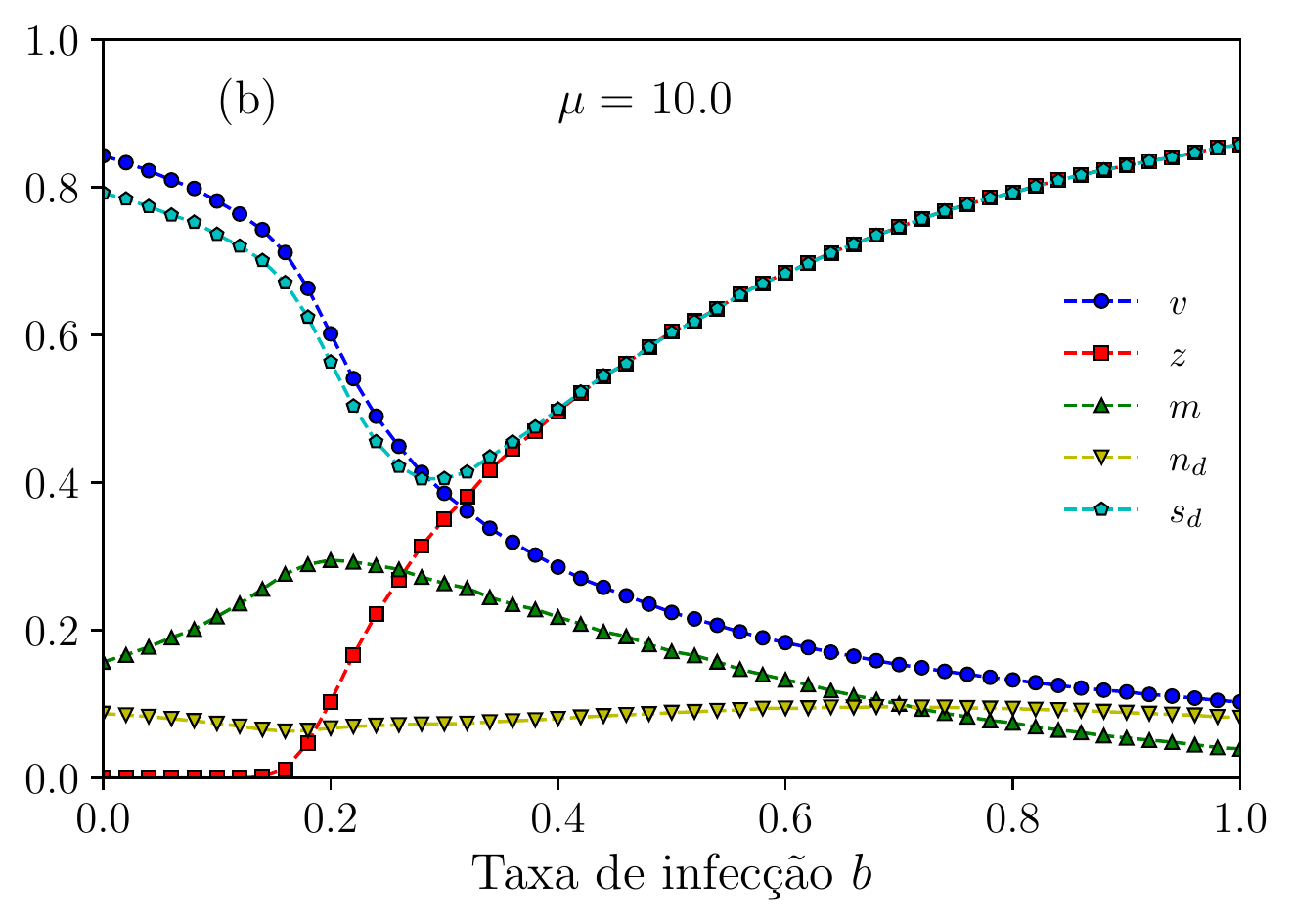}
\includegraphics[scale = 0.5]{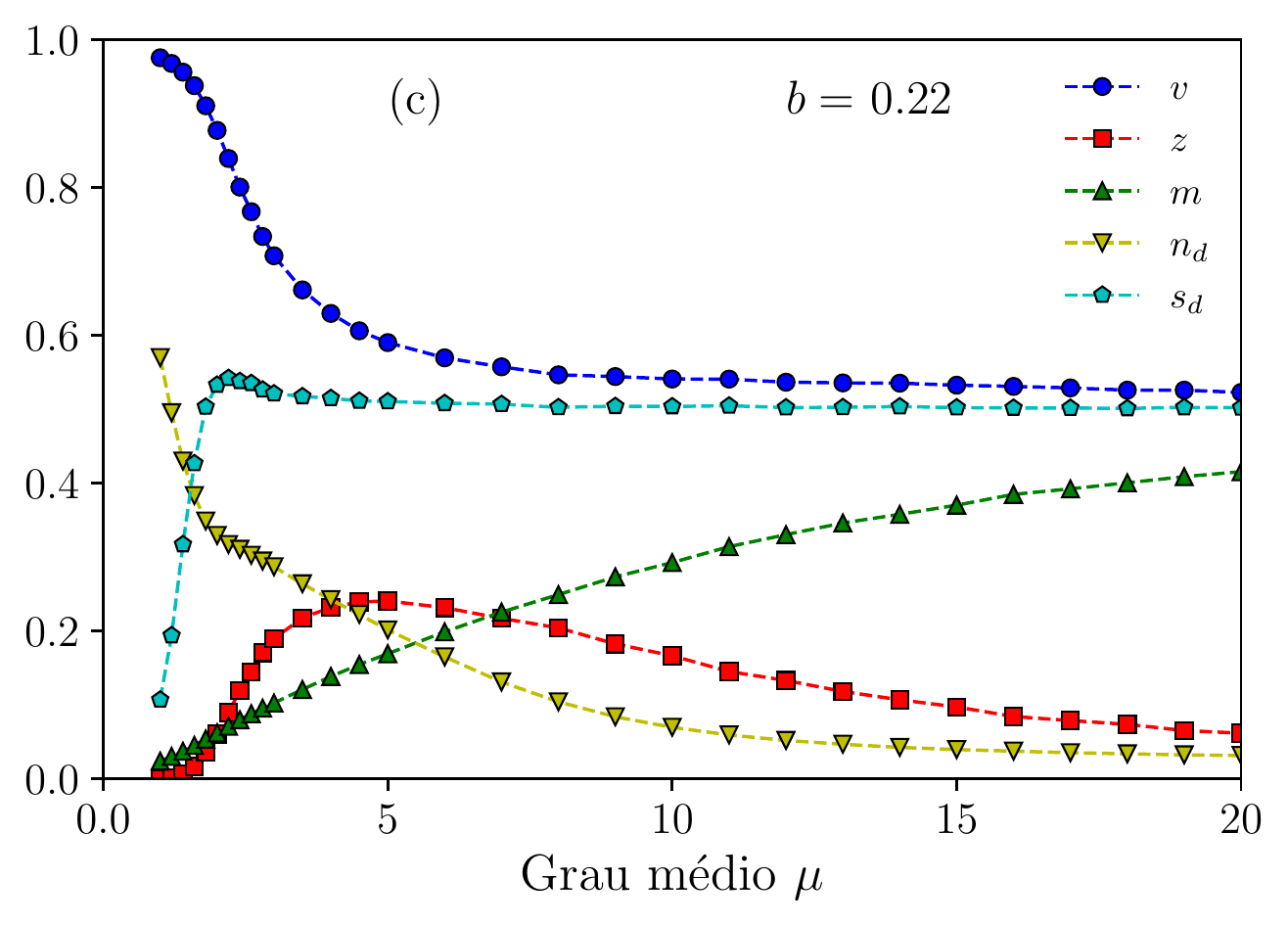}
\caption{Densidades $v$, $z$, $m$, $n_d$ e $s_d$. Parâmetros: $N = 16^2=256$, $z_0=1/2$, $Q=500$, $c=0.1$, $f=0.1$, $S_1=100$ e $S_2=50$. (a) Em função de $b$ com $\mu = 2.0$. (b) Em função de $b$ com $\mu = 10.0$. (c) Em função de $\mu$ com $b=0.22$.}
\label{figura6}
\end{center}
\end{figure}

\subsection{Domínios}

Os domínios também trazem informações que contribuem para o entendimento das fases do sistema. Os domínios são limitados pela topologia da rede e são definidos em última instância pela dinâmica do modelo. Como as unidades são as mesmas, podemos fazer uma comparação numérica entre os domínios e as densidades. Por exemplo: se o maior domínio $s_d$ for igual a densidade de vivos (ou mortos ou zumbis), podemos concluir que todos os vivos estão no maior domínio. Da mesma forma se por exemplo $z>s_d$, podemos concluir que esse maior domínio só contém zumbis.

A figura \ref{figura6}(a) apresenta as variações das densidades de indivíduos e de domínios em função de $b$ para $\mu = 2.0$. De imediato é possível notar a grande variação dos domínios, especialmente no intervalo $0.2<b<0.4$. Na fase absorvente ($b<0.2$) o maior domínio é composto por vivos ($v>s_d$) mas sofre uma queda brusca na transição de fase $b \sim 0.2$. Até $b = 0.6$ não é possível identificar o tipo de indivíduo que compõe o maior domínio, já que $v$ e $z$ são maiores que $s_d$. A medida que o maior domínio cai o número de domínios $n_d$ aumenta. Para $b$ grande, maior que 0.6, o maior domínio fica ocupado todo por zumbis pois $z>s_d$. Já na figura \ref{figura6}(b) temos as mesmas densidades mas para $\mu=10.0$. A primeira diferença é que devido ao grau médio maior há um menor número de componentes $n_d$. De fato, aumentando a conectividade as componentes aumentam de tamanho possibilitando os domínios aumentarem também. Na fase absorvente quase a totalidade dos vivos estão no maior domínio: $v$ um pouco maior que $s_d$. Já na fase ativa a partir de $b \sim 0.4$ todos os zumbis estão na maior componente: $z = s_d$.

Na figura \ref{figura6}(c) está mostrado os mesmos parâmetros porém em função do grau médio $\mu$ para $b=0.22$. O interessante neste resultado é que o tamanho do maior domínio $s_d$ é aproximadamente constante no intervalo $3<\mu < 20$ apesar do número de domínios $n_d$ diminuir consideravelmente. Isso significa que os domínios menores estão se juntando entre si enquanto o maior domínio se mantém inalterado. Como $v>s_d$ podemos afirmar que o maior domínio é formado inteiramente por indivíduos vivos. Além disso a densidade de mortos aumenta enquanto a de zumbis diminui. Isso nos permite concluir que os domínios pequenos estão se juntando pois indivíduos zumbis que separam esses domínios de mortos estão se convertendo em mortos, o que agrega os domínios.

\section{CONCLUSÕES}

Neste trabalho aplicamos uma variação do modelo epidêmico SIS para descrever um apocalipse zumbi no qual indivíduos vivos podem se tornar zumbis quando em contato com os mesmos. Além disso imaginamos que o estado zumbi seja temporário de forma que há uma chance constante de um zumbi se tornar novamente vivo. Essa volta faz com que o sistema apresente tanto a fase absorvente (todos os indivíduos são vivos) como também a fase ativa. Nesta última, no estado estacionário, a taxa de conversão de vivo para zumbi é igual a taxa contrária, de forma que ambas as densidades fiquem constantes ao longo do tempo. Além desses estados, adicionamos também um terceiro estado (morto) que pode ocorrer quando um zumbi entra em contato com um exterminador, que perfaz um pequeno grupo dos indivíduos vivos. Utilizamos como padrão de comunicação uma rede do tipo Erdös-Rényi (RER) e analisamos a influência do grau médio desta no diagrama de fases do sistema. Observamos que um baixo valor deste parâmetro dificulta o sistema a atingir a fase ativa, favorecendo o aumento da população dos indivíduos vivos. Além disso há um máximo na densidade de mortos para uma baixa taxa de infecção e alto valor de grau médio, já que assim a chance de um exterminador matar um zumbi aumenta. Utilizamos também a distribuição de domínios e componentes da rede para descrever em detalhes a população no estado estacionário.

Este estudo mostra que a topologia de uma rede complexa pode influenciar a dinâmica de um modelo, alterando suas propriedades. Como perspectiva, pode-se acoplar a dinâmica com a topologia de forma que um altera o outro criando tanto uma transição de fases no modelo quanto na topologia da rede.

\section{\textit{Agradecimentos}}

G. V. Sousa agradece o suporte financeiro de Rafael Grisotto e Souza e do Programa de Iniciação Científica da Universidade Federal de Jataí. P. F. Gomes agradece o auxílio financeiro da Fapeg (\url{http://www.fapeg.go.gov.br/}) e do CNPq (\url{www.cnpq.br}, processo 405508/2021-2). As simulações numéricas foram realizadas no LAMCAD / UFG (\url{https://lamcad.ufg.br/}).


\begin{thebibliography}{}

\bibitem[Grassberger, 1983]{Grassberger1983}
Grassberger, P. (1983).
\newblock Critical behavior of the general epidemic process and dynamical
  percolation.
\newblock {\em Mathematical Biosciences}, 63:157--172.

\bibitem[Hethcote, 2000]{Hethcote2000}
Hethcote, H.~W. (2000).
\newblock The mathematics of infectious diseases.
\newblock {\em SIAM Review}, 42(4):599--653.

\bibitem[Miranda et~al., 2020]{Miranda2020}
Miranda, L. H.~F., Ribeiro, B.~V., Rocha, P. M.~M., Santos, D. D.~A., and Sena,
  N. C.~d. (2020).
\newblock Surviving the zombie apocalypse: A population dynamics based
  approach.
\newblock {\em Revista Brasileira de Ensino de Física}, 42:1806--9126.

\bibitem[Reia and Fontanari, 2022]{Reia2022}
Reia, S.~M. and Fontanari, J.~F. (2022).
\newblock Long-term scientific impact revisited.
\newblock {\em Eur. Phys. J. Plus}, 137:161.

\bibitem[Barth\'{e}lemy et~al., 2005]{Barthelemy2005}
Barth\'{e}lemy, M., Barrat, A., Pastor-Satorras, R., and Vespignani, A. (2005).
\newblock Dynamical patterns of epidemic outbreaks in complex heterogeneous
  networks.
\newblock {\em Journal of Theoretical Biology}, 235:275--288.

\bibitem[da~Silva and Fernandes, 2015]{Silva2015}
da~Silva, R. and Fernandes, H.~A. (2015).
\newblock A study of the influence of the mobility on the phase transitions of
  the synchronous sir model.
\newblock {\em Journal of Statistical Mechanics: Theory and Experiment},
  2015(6):P06011.

\bibitem[Hinrichsen, 2000]{Hinrichsen2000}
Hinrichsen, H. (2000).
\newblock Non-equilibrium critical phenomena and phase transitions into
  absorbing states.
\newblock {\em Advances in Physics}, 49:815--958.

\bibitem[Vilela et~al., 2020]{Vilela2020}
Vilela, E.~B., Fernandes, H.~A., Costa, F. L.~P., and Gomes, P.~F. (2020).
\newblock Phase diagrams of the ziff–gulari–barshad model on random
  networks.
\newblock {\em Journal of Computational Chemistry}, 41:1964–1972.

\bibitem[Newman, 2010]{Newman2010}
Newman, M. E.~J. (2010).
\newblock {\em Networks: an introduction}.
\newblock Oxford University Press, Oxford.

\bibitem[Juh{\'a}sz et~al., 2015]{Juhasz2015}
Juh{\'a}sz, R., Kov{\'a}cs, I.~A., and Igl{\'o}i, F. (2015).
\newblock Long-range epidemic spreading in a random environment.
\newblock {\em Physical Review E}, 91(3):032815.

\bibitem[Gomes et~al., 2019]{Gomes2019}
Gomes, P.~F., Reia, S.~M., Rodrigues, F.~A., and Fontanari, J.~F. (2019).
\newblock Mobility helps problem-solving systems to avoid groupthink.
\newblock {\em Physical Review E}, 99:032301.

\bibitem[Erd\"{o}s and R\'{e}nyi, 1959]{Erdos1959}
Erd\"{o}s, P. and R\'{e}nyi, A. (1959).
\newblock On random graphs.
\newblock {\em Publicationes Mathematicae}, 6:290--297.

\bibitem[Erd\"{o}s and R\'{e}nyi, 1960]{Erdos1960}
Erd\"{o}s, P. and R\'{e}nyi, A. (1960).
\newblock On the evolution of random graphs.
\newblock {\em Publications of the Mathematical Institute of the Hungarian
  Academy of Sciences}, 5:17--61.
  
\bibitem[Solomonoff and Rapoport, 1951]{Solomonoff1561}
Solomonoff, R. and Rapoport, A. (1951).
\newblock Connectivity of random nets.
\newblock {\em B. Math. Biophys.}, 13:107--117.

\bibitem[Barab\'{a}si, 2016]{Barabasi2016}
Barab\'{a}si, A.-L. (2016).
\newblock {\em Network Science}.
\newblock Cambridge University Press, Glasgow.

\bibitem[Reia et~al., 2019]{Reia2019a}
Reia, S.~M., Gomes, P.~F., and Fontanari, J.~F. (2019).
\newblock Policies for allocation of information in task-oriented groups:
  elitism and egalitarianism outperform welfarism.
\newblock {\em The European Physical Journal B}, 92:205.

\bibitem[Reia et~al., 2020]{Reia2020a}
Reia, S.~M., Gomes, P.~F., and Fontanari, J.~F. (2020).
\newblock Comfort-driven mobility produces spatial fragmentation in axelrod’s
  model.
\newblock {\em Journal of Statistical Mechanics}, 033402:033402.

\bibitem[Gomes et~al., 2022]{Gomes2022}
Gomes, P.~F., Fernandes, H.~A., and Costa, A.~A. (2022).
\newblock Topological transition in a coupled dynamics in random networks.
\newblock {\em Physica A}, 597:127269.

\bibitem[Axelrod, 1997]{Axelrod1997}
Axelrod, R. (1997).
\newblock The dissemination of culture: a model with local convergence and
  global polarization.
\newblock {\em Journal of Conflict Resolution}, 41:203.



\bibitem[Klemm et~al., 2005]{Klemm2005}
Klemm, K., Egu\'iluz, V.~M., Toral, R., and Miguel, M.~S. (2005).
\newblock Globalization, polarization and cultural drift.
\newblock {\em Journal of Economic Dynamics and Control}, 29:321.

\bibitem[Reia and Fontanari, 2016]{Reia2016}
Reia, S.~M. and Fontanari, J.~F. (2016).
\newblock Effect of long-range interactions on the phase transition of
  axelrod's model.
\newblock {\em Physical Review E}, 94:052149.

\bibitem[Saberi, 2015]{Saberi2015}
Saberi, A.~A. (2015).
\newblock Recent advances in percolation theory and its applications.
\newblock {\em Physics Reports}, 578:1--32.

\bibitem[Harris and \textit{et al}, 2020]{Harris2020}
Harris, C.~R. and \textit{et al} (2020).
\newblock Array programming with numpy.
\newblock {\em Nature}, 585:357–362.

\bibitem[Hunter, 2007]{Hunter2007}
Hunter, J.~D. (2007).
\newblock Matplotlib: A 2d graphics environment.
\newblock {\em Computing in Science \& Engineering}, 9:90--95.

\bibitem[Hagberg et~al., 2008]{Hagberg2008}
Hagberg, A.~A., Schult, D.~A., and Swart, P.~J. (2008).
\newblock Exploring network structure, dynamics and function using networkx.
\newblock In {\em Proceedings of the 7th Python in Science Conference
  (SciPy2008)}, volume 445, page 11–15.

\bibitem[McKinney, 2010]{McKinney2010}
McKinney, W. (2010).
\newblock Data structures for statistical computing in python.
\newblock In {\em Proceedings of the 9th Python in Science Conference}, volume
  445, pages 51--56.


\end{thebibliography}
\end{document}